\documentstyle[pra,aps,psfig]{revtex}
\newcommand{\beq}{\begin{equation}}
\newcommand{\eeq}{\end{equation}}
\newcommand{\beqa}{\begin{eqnarray}}
\newcommand{\eeqa}{\end{eqnarray}}
\newcommand{\ba}{\begin{array}}
\newcommand{\ea}{\end{array}}

\begin{document}
\draft

\twocolumn[\hsize\textwidth\columnwidth\hsize\csname
@twocolumnfalse\endcsname

\widetext 
\title{Beyond Mean-Field Theory for Attractive Bosons \\ 
under Transverse Harmonic Confinement} 
\author{Luca Salasnich} 
\address{CNR-INFM and CNISM, Unit\`a di Milano Universit\`a, 
c/o Dipartimento di Fisica, Universit\`a di Milano, \\
Via Celoria 16, 20133 Milano, Italy}

\maketitle

\begin{abstract} 
We study a dilute gas of attractive bosons confined in 
a harmonic cylinder, i.e. under cylindric confinement due 
to a transverse harmonic potential. 
We introduce a many-body wave function which extends 
the Bethe ansatz proposed by McGuire (J. Math. Phys. {\bf 5}, 622 (1964)) 
by including a variational transverse Gaussian shape. 
We investigate the ground state properties of the system 
comparing them with the ones of the one-dimensional (1D) attractive Bose gas. 
We find that the gas becomes ultra 1D as a consequence 
of the attractive interaction: the transverse width of the Bose gas 
reduces by increasing the number 
of particles up to a critical width below which there is 
the collapse of the cloud. In addition, we derive a simple analytical 
expression for the simmetry-breaking solitonic density profile 
of the ground-state, which generalize the one deduced by Calogero 
and Degasperis (Phys. Rev. A {\bf 11}, 265 (1975)). This 
bright-soliton analytical solution shows near the collapse 
small deviations with respect to the 3D mean-field numerical solution. 
Finally, we show that our variational Gauss-McGuire theory is always 
more accurate than the McGuire theory. In addition, we prove that 
for small numbers of particles the Gauss-McGuire theory is more 
reliable than the mean-field theory described by the 3D 
Gross-Pitaevskii equation. 
\end{abstract}


]

\narrowtext 

\section{Introduction} 

A three-dimensional (3D) dilute Bose gas is well described 
by the mean-field Gross-Pitaevskii theory \cite{gross,pitaevskii}, 
while in the 1D regime quantum fluctuations become important 
and an appropriate theoretical treatment requires beyond mean-field 
approaches \cite{astra}. Recently, we have investigated, for a bosonic 
cloud of atoms, the crossover from 3D to 1D 
induced by a strong harmonic confinement 
in the transverse cylindric radial direction 
\cite{sala1,sala2,sala3,sala4}. 
In particular, by using a generalized Lieb-Liniger 
theory \cite{lieb,sala4}, 
based on a variational treatment of the transverse width of the repulsive 
Bose gas, we have analyzed the transition from a 3D Bose-Einstein condensate 
to a 1D Tonks-Girardeau gas of impenetrable bosons. 
\par 
In this paper we consider the case of an attractive 
Bose gas under cylindric transverse harmonic confinement. We introduce 
a trial wave fuction that is a variational extension of the 1D exact Bethe 
ansatz \cite{bethe} proposed by McGuire \cite{mcguire} for 
the 1D gas of bosons with attractive contact interaction. 
We show that, contrary to the 1D theory where the transverse width 
is constant and equal to the characteristic length of harmonic 
confinement, our theory predicts that the transverse 
width of the gas reduces by increasing the interatomic strength 
up to a critical value for which there is the collapse of the system. 
We investigate also the solitonic axial density profile of the Bose gas 
with a fixed center of mass, comparing it to the 1D profile 
and also to the 3D mean-field Gross-Pitaevskii theory. 

\section{Transverse Gaussian ansatz} 

The Hamiltonian of a gas of $N$ interacting identical 
Bose atoms confined in the transverse cylindric radial direction 
with a harmonic potential of frequency $\omega_{\bot}$ 
is given by 
\beq
{\hat H} = \sum_{i=1}^N \left( -{1\over 2} \nabla_i^2
+ {1\over 2}(x_i^2+y_i^2) \right) 
+ \sum_{i<j=1}^N V({\bf r}_i,{\bf r}_j) \; , 
\eeq 
where $V({\bf r}_i,{\bf r}_j)$ is the inter-atomic potential. 
In the Hamiltonian we use scaled units: 
energies are in units of the energy 
$\hbar \omega_{\bot}$ of the transverse confinement 
and lengths in units of the characteristic harmonic 
length $a_{\bot}=({\hbar/(m\omega_{\bot})})^{1/2}$. 
\par 
The determination of the $N$-body wave function 
$\Psi({\bf r}_1,...,{\bf r}_N)$ that minimizes the energy 
\beq 
E = {\langle \Psi |{\hat H}| \Psi \rangle} = 
\int \Psi^* {\hat H}\Psi \; d^3{\bf r}_1 ... d^3{\bf r}_N 
\eeq
of the system is a very difficult task. Nevertheless, 
due to the symmetry of the problem, a variational trial wave 
function can be written in the form 
\beq 
\Psi({\bf r}_1,...,{\bf r}_N) = 
f(z_1,...,z_N) 
\prod_{i=1}^N { 
\exp{ \left(-{x_i^2+y_i^2\over 2 \sigma^2} \right) } 
\over \pi^{1/2} \sigma} \; , 
\eeq 
where $\sigma$ is a variational parameter of the Gaussian transverse 
wave function \cite{perez}, that gives the effective 
transverse length of the Bose gas. 
In addition, by considering a dilute gas with a mean particle spacing 
much larger than the interaction radius we set 
\beq 
V({\bf r}_i,{\bf r}_j)= \Gamma \; 
\delta^{(3)}({\bf r}_i-{\bf r}_j) \; , 
\eeq 
where $\Gamma= 4\pi a_s/a_{\bot}$, with $a_s$ 
the s-wave scattering length of the inter-atomic potential. 
This pseudo-potential gives the correct dilute gas limit 
and a well-posed variational problem by choosing a smooth trial 
wave function. Under transverse harmonic trapping a 
confinement-induced resonance has been predicted by Olshanii 
\cite{olshanii} at $|a_s|/a_{\bot}\simeq 1$, i.e.  
when the absolute value of 
the highest bound-state energy 
of a realistic inter-atomic potential approaches the confining 
transverse energy. Therefore the pseudo-potential of Eq. (4) can be 
used in the range $|a_s|/a_{\bot} \ll 1$, 
a regime where the effects of confinement-induced resonance 
are negligible. 
\par 
By inserting Eq. (3) into Eq. (2), using Eq. (4) and integrating 
over $x$ and $y$, the total energy reads 
\beq 
E = E_z + E_{\bot} \; , 
\eeq
where the longitudinal axial energy is 
\beq 
E_z = 
\langle f |\sum_{i=1}^N  -{1\over 2} {\partial^2 \over \partial z_i^2}  
+ {\Gamma \over 2\pi \sigma^2 } 
\sum_{i<j=1}^N \delta(z_i - z_j) | f \rangle  \; . 
\eeq 
It is important to stress that the 1D Hamiltonian 
that appears in the previous expresssion is exactly the Hamiltonian 
studied by Lieb and Liniger \cite{lieb} 
for positive interaction strength ($\Gamma/(2\pi \sigma^2) >0$) 
and by Mc Guire \cite{mcguire} 
and by Calogero and De Gasperis \cite{calogero} 
for negative interaction strength ($\Gamma/(2\pi \sigma^2) <0$). 
The transverse radial energy is instead given by 
\beq 
E_{\bot} = {1\over 2} ({1\over \sigma^2}+\sigma^2) N \; . 
\eeq 
As previously stressed, with the ansatz of Eq. (3) we have recently 
analyzed \cite{sala3} the repulsive case ($a_s>0$) by using the Lieb-Liniger 
exact result \cite{lieb} for the longitudinal axial energy. 
Here we consider the attractive case ($a_s<0$) and set 
${\Gamma / 2\pi \sigma^2}
= -\gamma /\sigma^2$ with $\gamma = 2|a_s|/a_{\bot}$. 

\section{Gauss-McGuire ansatz} 

McGuire \cite{mcguire} proposed the following Bethe ansatz \cite{bethe} 
for the axial many-body wave function 
\beq 
f(z_1,...z_N) = C_N \prod_{1\le i<j \le N} 
\exp{\left( -{1\over 2}{\gamma\over\sigma^2}|z_i - z_j| 
\right)} \; , 
\eeq 
where $C_N$ is the normalization constant 
and $\gamma /\sigma^2$ is the strength of the contact 
$\delta$ interaction in the 1D Hamiltonian of Eq. (6). 
According to the exact result of Mcguire \cite{mcguire} 
the axial energy $E_z$ of Eq. (6) reads 
\beq 
E_z = - {1\over 24} {\gamma^2\over\sigma^4} N (N^2 -1) \; . 
\eeq 
Using this expression the total energy $E$ of our 3D system 
can be rewritten as 
\beq 
E = -{1\over 24} {\gamma^2\over\sigma^4} (N^2 -1)N + 
{1\over 2} ({1\over \sigma^2}+\sigma^2)N \; . 
\eeq 
In the weak-coupling 1D limit, where the 
transverse width $\sigma \simeq 1$, one recovers the 
1D result obtained by McGuire \cite{mcguire} plus the constant transverse 
energy, that is $1$ in units of $\hbar\omega_{\bot}$. 
\par 
In our approach the energy depends on the variational parameter $\sigma$. 
The minimization of the energy $E$ with respect to $\sigma$ gives the equation 
\beq 
\sigma^6 - \sigma^2 + {1\over 6}\gamma^2 (N^2 -1) = 0 \; .   
\eeq
One easily finds that this algebric equation admits real solutions 
with $d^2 E/d\sigma^2>0$ if and only if 
\beq 
{1 \over 3^{1/4}} \le \sigma \le 1 \; . 
\eeq 
Below the value $\sigma =1/3^{1/4}$ there are no stable solutions 
and the ground state becomes 
the collapsed state, i.e. the configuration with $\sigma =0$ and 
energy $E=-\infty$. The critical strength, corresponding 
to $\sigma =1/3^{1/4}$, is 
\beq 
\gamma (N^2-1)^{1/2} 
= {2 \over 3^{1/4}} \simeq 1.52 \; . 
\eeq  

\begin{figure}
\centerline{\psfig{file=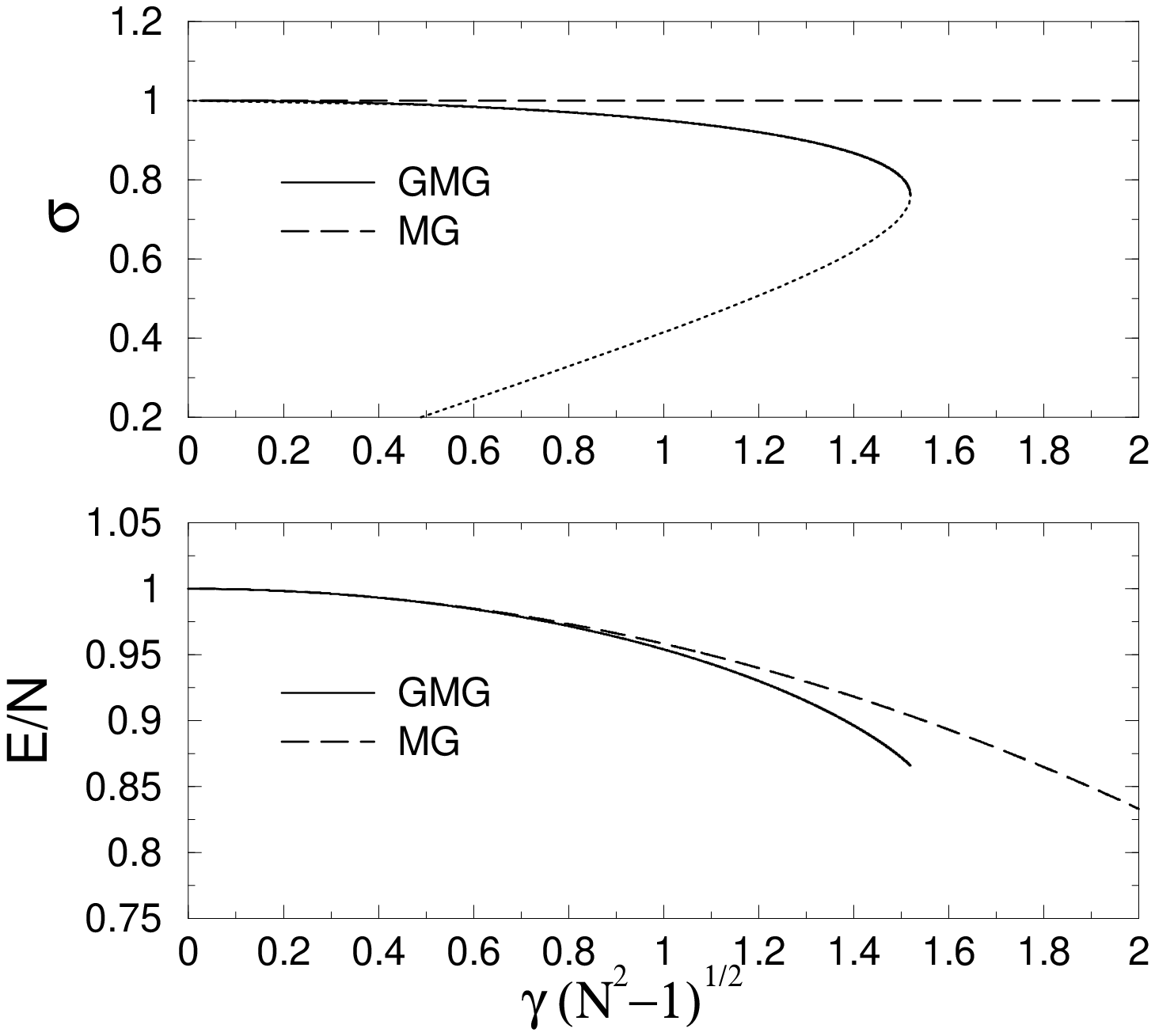,height=3.in,clip=}}
{FIG. 1. Comparison between our variational Gauss-McGuire 
theory (GMG) and the McGuire theory (MG) for the attractive 
Bose gas in a harmonic cylinder. Upper panel: transverse width $\sigma$. 
The unstable branch ($d^2E/d\sigma^2<0$) of Eq. (11) 
is shown as a dotted line. Lower panel: energy per particle $E/N$. 
$\gamma = 2|a_s|/a_{\bot}$ is the inter-atomic strength 
and $N$ is the number of particles. }
\end{figure}

\par 
In the upper panel of Fig. 1 we plot the transverse width $\sigma$ 
obtained from Eq. (11) as a function of $\gamma (N^2-1)^{1/2}$ (solid line). 
While in the 1D case $\sigma$ is constant (dashed line), 
our variational Gauss-McGuire (GMG) method shows that 
the attractive Bose gas is actually ultra 1D: 
the width $\sigma$ decreases by increasing the 
interaction strength up to the collapse. 
In the lower panel of Fig. 1 we plot the energy per particle $E/N$ 
as a function of $\gamma (N^2-1)^{1/2}$. 
We compare our variational energy given by Eqs. (10,11) 
with the McGuire (MG) energy given by Eq. (10) and $\sigma =1$. 
As expected the variational GMG energy 
is lower than the MG energy giving a better determination 
(upper bound) of the true ground state of the many-body system. 
\par 
It is important to observe that the axial McGuire wave function 
is invariant by a global translation of the positions of the particles. 
As shown by Calogero and Degasperis \cite{calogero}, and also by 
Castin and Herzog \cite{castin}, the wave function of Eq. (8) can be 
normalized by imposing that 
its center of mass $z_{cm}$ is fixed, namely 
\beq 
\int \delta(z-z_{cm}) |f|^2 \; dz_1 ... dz_N = N \; . 
\eeq
The density of particles with respect to $z_{cm}$ is then given by 
\beq 
\rho(z) = \int \delta(z_{cm})  
\delta(z_1 - z) |f|^2 \; dz_1 \; ... \; dz_N \; . 
\eeq
Following Calogero and Degasperis \cite{calogero} 
we immediately find 
\beq 
\rho(z) = {\gamma \over \sigma^2} 
\sum_{l=1}^{N-1} {(-1)^{l+1} l (N!)^2 
\exp{\left( - \gamma l N |z|/\sigma^2 \right)} 
\over (N+l-1)! (N-l-1)!} \; . 
\eeq 
For $N\gg1$ this formula can be approximated by 
\beq 
\rho(z) = {\gamma N^2\over 4\sigma^2} 
sech^2\left({\gamma N\over 2 \sigma^2}z\right) \; , 
\eeq
that is the first term in a $1/N$ expansion of the previous expression. 
Obviously, only for $\sigma = 1$ the Eq. (16) gives the 
1D solitonic profile predicted by Calogero and Degasperis 
\cite{calogero}. In our variational GMG scheme $\sigma$ is not constant 
and must be determined using the Eq. (11). 

\section{Comparison with the 3D GPE}

Here we compare the GMG wave function with the fully 3D Hartree 
wave function, from which one obtains the 3D Gross-Pitaevskii 
equation (GPE). 
The 3D Hartree approximation is obtained by setting 
\beq 
\Psi({\bf r}_1,...,{\bf r}_N) = N^{1/2} 
\prod_{i=1}^N \psi({\bf r}_i) \; . 
\eeq 
Inserting this many-body wave function into Eq. (2) with Eq. (4), 
after integration one finds the Gross-Pitaevskii energy functional 
$$
E[\psi({\bf r})] = N \int \psi^* 
\left[ -{1\over 2} \nabla^2 + {1\over 2} (x^2 + y^2) \right. 
$$
\beq 
\left. 
- {1\over 2} 2\pi \gamma (N-1)  |\psi |^2 
\right] \psi \; d^3{\bf r} \; . 
\eeq
By minimizing this energy functional with respect to $\psi({\bf r})$ 
with the constraint of the normalization 
$\int |\psi |^2 d^3{\bf r}= 1$ 
one finds the familiar mean-field 3D GPE given by 
\beq 
\left[ -{1\over 2} \nabla^2 + {1\over 2}(x^2 + y^2) 
- 2 \pi \gamma (N-1) |\psi |^2 \right] \psi 
= \mu \; \psi \; ,  
\eeq 
where $\mu$ is the chemical potential fixed by the normalization. 
We solve the stationary GPE by using a finite-difference 
Crank-Nicolson predictor-corrector 
algorithm with imaginary time \cite{sala5}. 

\begin{figure}
\centerline{\psfig{file=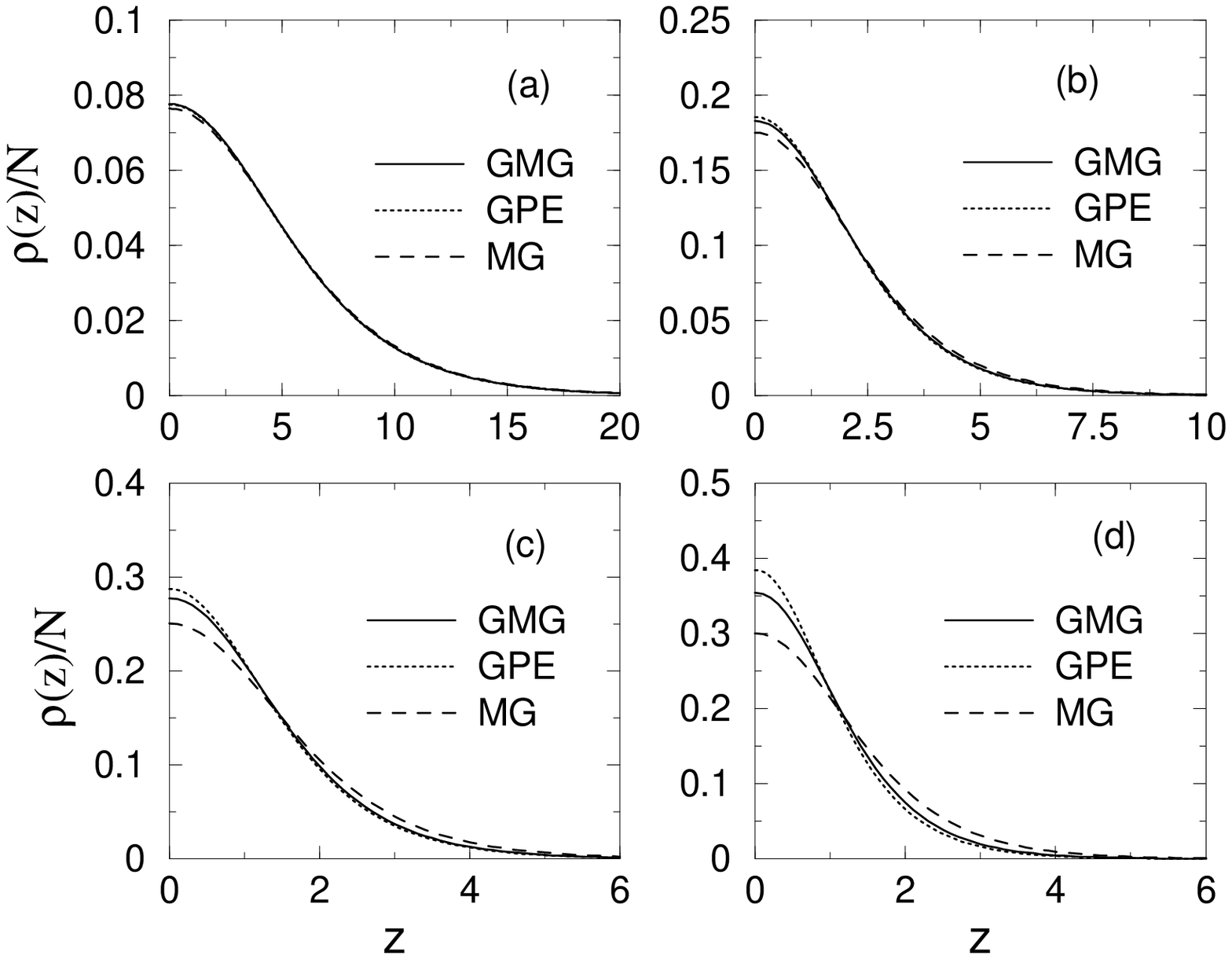,height=2.7in,clip=}}
{FIG. 2. Density profile $\rho(z)$ of the bright soliton
solution obtained with three different theories:
Gauss-McGuire (GMG), McGuire (MG) and the Gross-Pitaevskii equation 
(GPE).Four values of the interaction strength:
(a) $\gamma (N-1) =0.7$; (b) $\gamma (N-1)=0.9$; (c) $\gamma (N-1) = 1$; 
(d) $\gamma (N-1) = 1.2$. $N$ is the number of particles and
$\gamma =0.001$.}
\end{figure}

\par 
In Fig. 2 we plot the solitonic density profile of the GPE 
and compare it with the profile of GMG theory, 
i.e. Eq. (16) with $\sigma$ given by Eq. (11). 
We insert also the solitonic profile of the MG theory, 
i.e. Eq. (16) with $\sigma =1$. 
Up to the collapse our GMG 
solitionic profile is close to the GPE one; 
there are instead relevant deviations with respect 
to the MG soliton (which does not collapse). 
Our numerical integration of the GPE gives 
the collapse for $\gamma (N-1)\simeq 1.35$ in agreement 
with previous computations \cite{gammal} and not too far from 
the analytical prediction of Eq. (13). 
The GPE critical strength is very close to the analytical 
prediction $\gamma (N-1) = 4/3 \simeq 1.33$ of the non-polynomial
Schrodinger equation (NPSE) \cite{sala1,sala2}. 
However the NPSE is not an exact variational equation: 
it was obtained from the GPE by neglecting some space 
derivatives (for details see \cite{sala1} and \cite{russi}). 
\par 
The results of Fig. 2 are obtained by setting 
$\gamma=0.001$. In Fig. 3 we use the same value of $\gamma$, 
that is relevant for the experiments performed with Bose-Einstein 
condensates of $^7$Li atoms \cite{esperimento1,esperimento2}. 
Fig. 3 shows the energy per particle $E/N$, the chemical potential 
$\mu$, the density per particle $\rho(0)/N$, and also 
the transverse width $\sigma$ of the soliton 
as a function of the number $N$ of bosons. 
The GMG energy (solid line) 
is always close to the GPE energy (dotted line) 
and near the collapse the energy difference slightly increases. 
Chemical potentials and densities have a similar behavior. 
Fig. 3 shows that for a large number $N$ of bosons the MG 
theory (dashed lines) differs with respect to the 
other two theories: for $N >750$ deviations are clearly 
visible. 

\begin{figure}
\centerline{\psfig{file=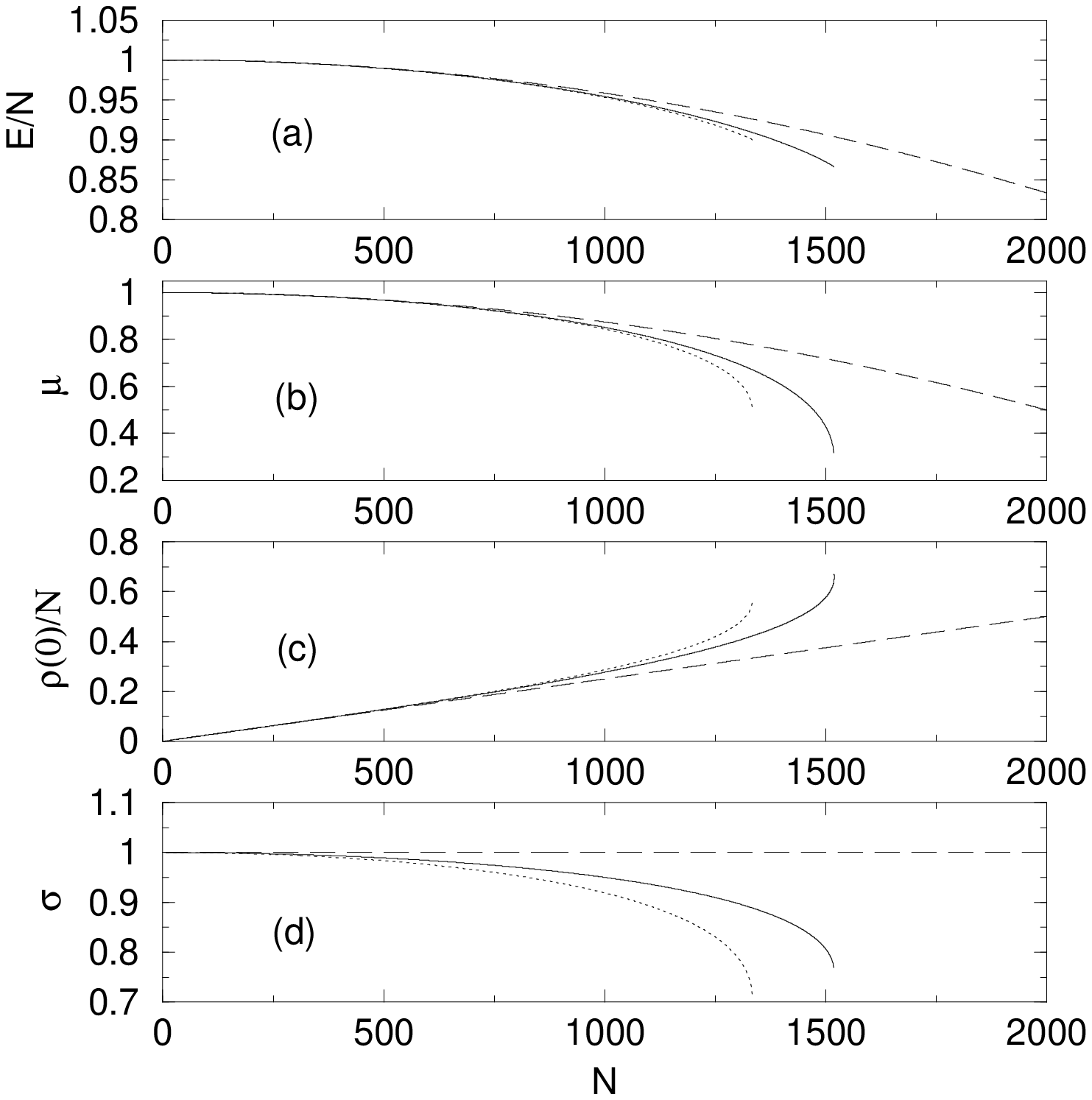,height=3.6in,clip=}}
{FIG. 3. Bright soliton properties. 
Comparison among different theories: 
Gauss-McGuire (solid line), McGuire (dashed line) 
and the Gross-Pitaevskii equation (dotted line). 
(a) Energy per particle $E/N$. 
(b) Chemical potential $\mu$. 
(c) Axial density per particle $\rho(0)/N$ 
at the origin. (d) Transverse width $\sigma$ at $z=0$. 
$N$ is the number of particles and $\gamma =0.001$.}
\end{figure}

The panel (d) of Fig. 3 shows that the transverse width $\sigma$ 
decreases by increasing $N$ for both the GPE and 
the GMG theory; however $\sigma$ stays always of the order 
of one up to the collapse. Note that the transverse 
width $\sigma$ obtained by using the GPE depends on the axial 
coordinate $z$. In the plot we choose $\sigma$ at $z=0$, 
that is the lower value of the transverse width. 
\par 
The variational principle says that the 
lowest energy indicates the most accurate solution. 
The variational principle applies when different variational 
solutions for the same Hamiltonian are compared. 
In our case, the Hamiltonian is given by Eq. (1) with Eq. (4) 
and the three variational solutions are: the GMG one 
given by Eq. (3) and Eq. (8), the MG one given by 
Eq. (3) with $\sigma=1$ and Eq. (8), and the GPE one 
given by Eq. (19). For $N=1$ the three theories 
give the same value of energy: $E/N=1$. The top panel of Fig. 3 
shows that for a large $N$ the GPE gives the lowest energy. 
We expect that for a small $N$ 
the GMG theory and the MG theory are more reliable 
than the GPE. To verify this prediction we choose a larger 
value of $\gamma$, namely $\gamma=0.01$, and calculate 
numerically the GPE energy of Eq. (19) for all integer values 
of $N$ up the the collapse. The results are shown in Fig. 4 
where we plot the energies for increasing values of $N$. 

\begin{figure}
\centerline{\psfig{file=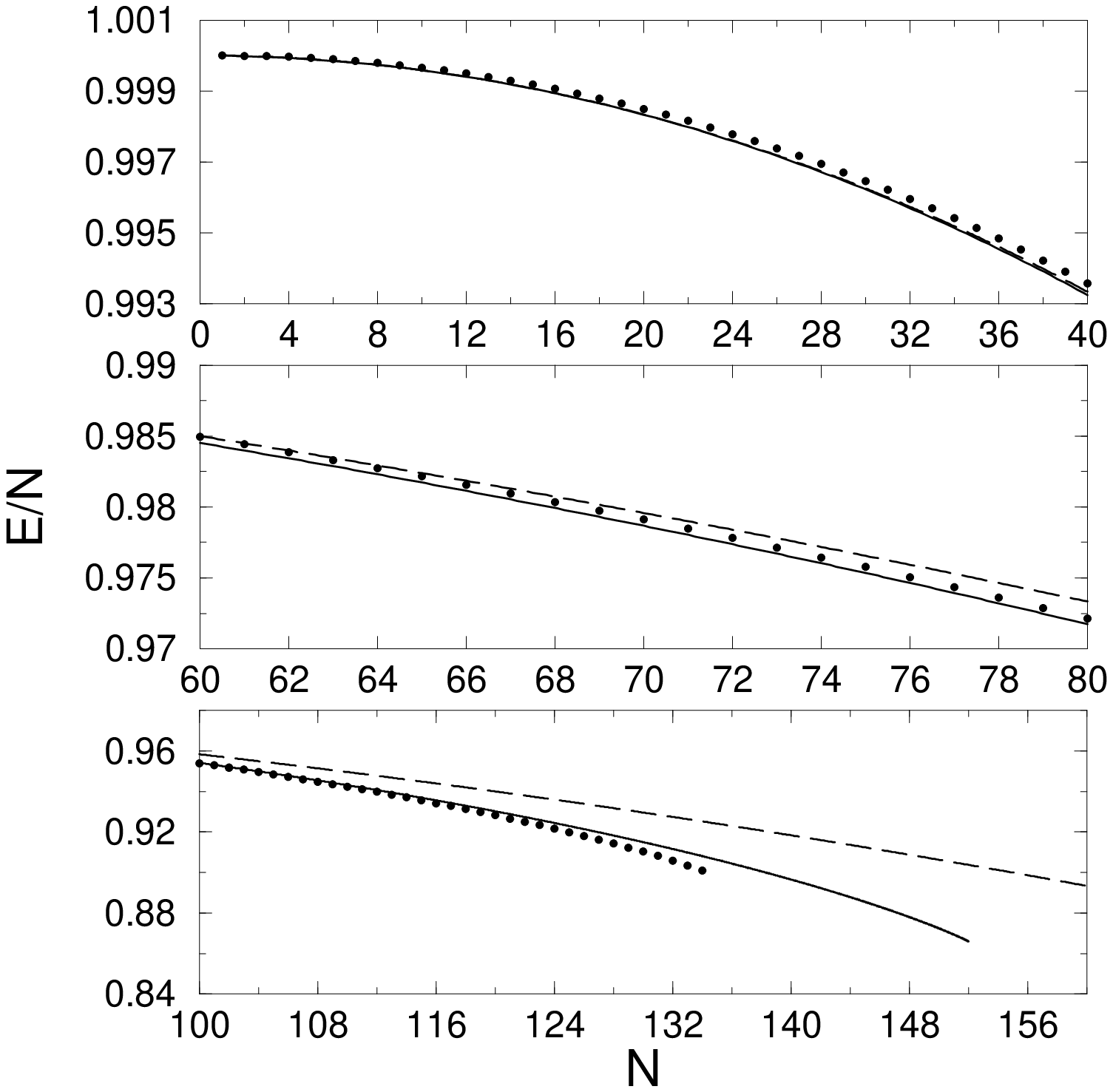,height=3.4in,clip=}}
{FIG. 4. Energy per particle $E/N$ as a function 
of the number $N$ of bosons. 
Comparison among different theories: 
Gauss-McGuire (solid line), McGuire (dashed line) 
and the Gross-Pitaevskii equation (filled circles). 
Here $\gamma =0.01$.}
\end{figure}

For $1<N<40$ the GMG energy (solid line) and 
the MG energy (dashed line) are indistinguishable but 
both lower than the GPE energy 
(filled circles). Around $N=60$ the GPE energy becomes lower then 
the MG energy but still higher than the GMG energy. 
Only around $N=100$ the GPE energy becomes smaller than the 
GMG energy. At $N=136$ there is the collapse of the 
GPE solitonic solution while the GMG soliton 
collapses at $N=153$. Just before the collapse of the 
GPE solution, i.e. for $N=135$, the relative difference $\Delta_R$ 
between GPE energy and GMG energy is $0.79\%$, 
while $\Delta_R$ between GPE energy and 
MG energy is $2.68\%$ (see also the upper panel of Fig. 5). 

\begin{figure}
\centerline{\psfig{file=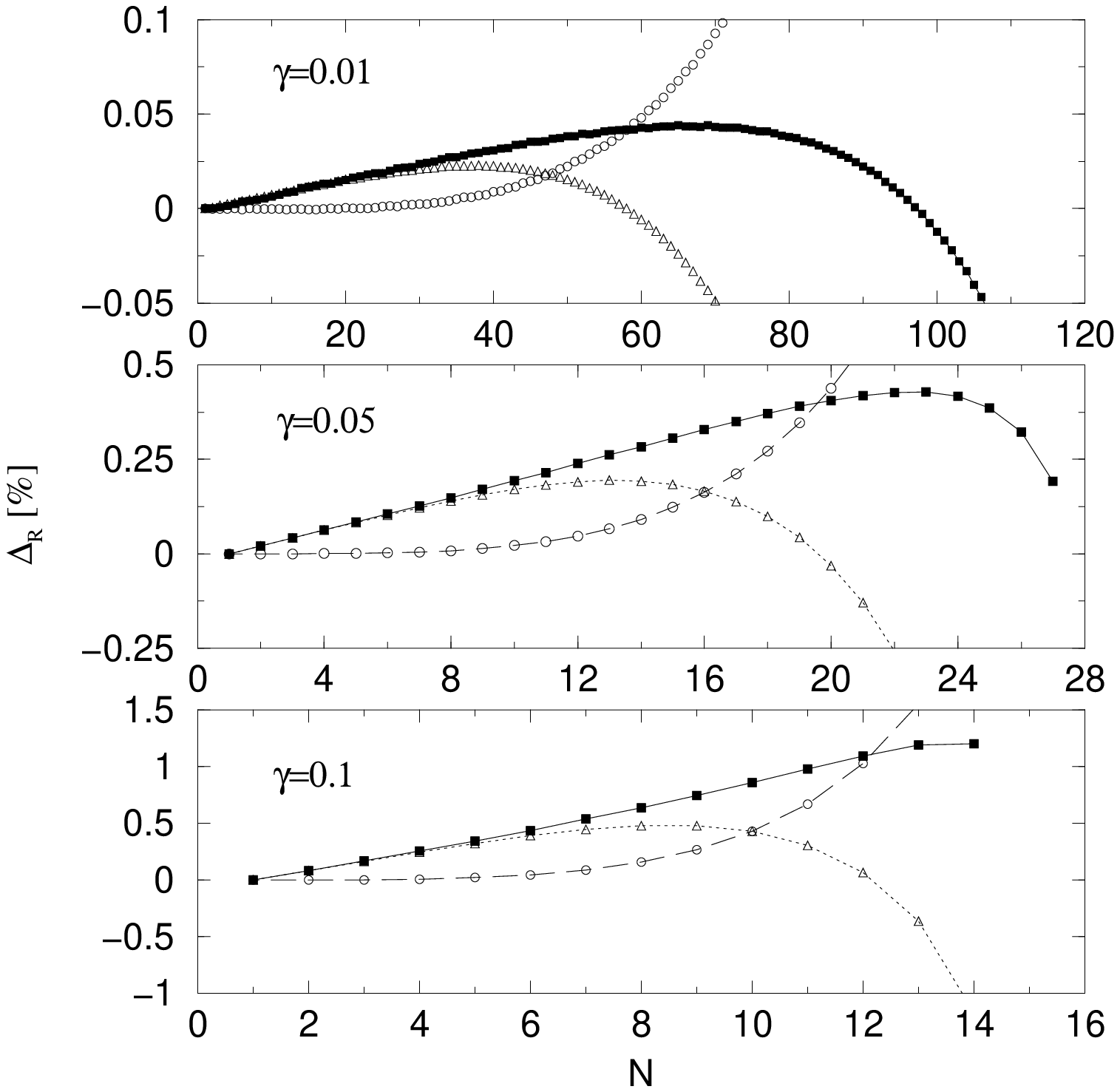,height=3.2in,clip=}}
{FIG. 5. Energetic relative difference $\Delta_R$ (percentual) 
as a function of the number $N$ of bosons. 
Circles: $\Delta_R$ between McGuire and Gauss-McGuire.
Filled squares: $\Delta_R$ between Gross-Pitaevskii equation and 
Gauss-McGuire. Triangles: $\Delta_R$ between Gross-Pitaevskii 
equation and McGuire.} 
\end{figure}

Similar results are obtained by setting $\gamma =0.001$. 
Here the energetic relative difference $\Delta_R$ 
between GPE and MG theory changes sign 
around $N=270$. Instead $\Delta_R$ between GPE and 
GMG theory changes sign around $N=480$. 
Just before the collapse of the GPE solution, i.e for $N=1334$, 
$\Delta_R$ is $2.86\%$ and $1.00\%$ respectively. 
\par 
To better study beyond mean-field effects we perform numerical 
calculations for various values of $\gamma$ 
in the interval $[0.01,0.1]$. The results are reported 
in Fig. 5, where we plot the energetic relative difference $\Delta_R$ 
for three values of the inter-atomic strength $\gamma$. 
The figure shows that the energy of the GMG theory is always 
smaller than the energy of the MG theory; in fact, their 
relative difference $\Delta_R$ (curves with circles) 
is always positive. The behavior of the curves 
in the three panels reveals 
that by growing $\gamma$ the range of $N$ reduces and 
the range of $\Delta_R$ increases. The curves with triangles 
(GPE energy minus GMG energy) and also the curves with 
filled squares (GPE energy minus MG energy) 
show that by fixing the number $N$ of particles and 
increasing the strength $\gamma$, the beyond mean-field effects 
become more important. Fig. 5 shows that for 
$\gamma =0.05$ and $\gamma =0.1$ the GMG theory gives 
a lower energy than the GPE up to the collapse. 
We find that the critical strength $\gamma_s$ beyond which 
the GMG theory is better than the GPE for all $N$ is 
$\gamma_s=0.044$; in this regime the critical number $N_c$ 
of particles for the collapse predicted by the GMG theory 
is more reliable than the GPE one. 

\section{Conclusions} 
We have introduced a beyond mean-field many-body 
wave function which describes dilute attractive bosons 
under transverse harmonic confinement. This variational 
approach, that we have called Gauss-McGuire theory, gives simple 
analytical formulas for the ground-state energy and the solitonic 
density profile. These formulas are in good agreement 
with the numerical results of the 3D mean-field theory. 
By comparing the ground-state energies, we have verified that 
for small numbers of particles the Gauss-McGuire theory 
is better than the 3D mean-field theory while 
the 3D mean-field theory is more reliable 
for a large number of particles and a small 
inter-atomic strength. 

\section*{Acknowledgements} 

We thank A. Parola, L. Reatto, Y. Castin and A. Recati 
for useful discussions.


\begin{thebibliography}{99}

\bibitem{gross} E.P. Gross, Nuovo Cimento {\bf 20}, 454 (1961). 

\bibitem{pitaevskii} L.P. Pitaevskii, Sov. Phys. JETP {\bf 13}, 451 (1961). 

\bibitem{astra} See, for instance, G.E. Astrakharchik, D. Blume, S. Giorgini, 
and B.E. Granger, J. Phys. B: At. Mol. Opt. Phys. {\bf 37}, S205 (2004). 

\bibitem{sala1} L. Salasnich, Laser Phys. {\bf 12}, 198 (2002); 
L. Salasnich, A. Parola, and L. Reatto, 
Phys. Rev. A {\bf 65}, 043614 (2002).  

\bibitem{sala2} L. Salasnich, A. Parola, and L. Reatto, 
Phys. Rev. A {\bf 66}, 043603 (2002). 

\bibitem{sala3} L. Salasnich, A. Parola, and L. Reatto, 
Phys. Rev. Lett. {\bf 91}, 080405 (2003); 
L. Salasnich, A. Parola, and L. Reatto, 
Phys. Rev. A {\bf 69}, 045601 (2004).

\bibitem{sala4} L. Salasnich, A. Parola, and L. Reatto, 
Phys. Rev. A {\bf 70}, 013606 (2004); 
L. Salasnich, A. Parola, and L. Reatto,
Phys. Rev. A {\bf 72}, 025602 (2005).  

\bibitem{lieb} E.H. Lieb and W. Liniger, Phys. Rev. {\bf 130}, 1605 (1963).  

\bibitem{olshanii} M. Olshanii, Phys. Rev. Lett. {\bf 81}, 938 (1998). 

\bibitem{bethe} H.A. Bethe, Z. Physik {\bf 71}, 205 (1931). 

\bibitem{mcguire} J.B. McGuire, J. Math. Phys. {\bf 5}, 622 (1964). 

\bibitem{calogero} F. Calogero and A. Degasperis, 
Phys. Rev. A {\bf 11}, 265 (1975). 

\bibitem{perez} V.M. Perez-Garcia, H. Machinel, and H. Herrero, 
Phys. Rev. A {\bf 57}, 3837 (1998); L. Salasnich, 
Int. J. Mod. Phys. B {\bf 14}, 1 (2000). 

\bibitem{castin} Y. Castin and C. Herzog, Comptes Rendus de 
l'Academie des Sciences de Paris, 
tome 2, serie IV,  419-443 (2001). 

\bibitem{sala5} E. Cerboneschi, R. Mannella, E. Arimondo, and L. Salasnich, 
Phys. Lett. A {\bf 249}, 495 (1998); L. Salasnich, A. Parola, 
and L. Reatto, Phys. Rev. A {\bf 64}, 023601 (2001). 

\bibitem{gammal} A. Gammal, L. Tomio and T. Frederico,
Phys. Rev. A {\bf 66}, 043619 (2002).

\bibitem{russi} A.M. Kamchatnov and V.S. Shchesnovich, 
Phys. Rev. A {\bf 70}, 023604 (2004). 

\bibitem{esperimento1} 
K.E. Strecker, G.B. Partridge, A.G. Truscott, and 
R.G. Hulet, Nature {\bf 417}, 150 (2002).  

\bibitem{esperimento2} 
L. Khaykovich, F. Schreck, G. Ferrari, T. Bourdel, 
J. Cubizolles, L.D. Carr, Y. Castin, and 
C. Salomon, Science {\bf 296}, 1290 (2002).  


\end{thebibliography}
\end{document}